\pdfoutput=1
\documentclass[aps,prl,reprint,twocolumn,superscriptaddress,amsmath,amssymb]{revtex4-2}

\usepackage{graphicx}
\usepackage{bm}
\usepackage{pgfplots}
\pgfplotsset{compat=1.18}

\newcommand{\kB}{k_{\mathrm{B}}}
\newcommand{\Eop}{E_{\mathrm{op}}}
\newcommand{\avg}[1]{\left\langle #1 \right\rangle}
\newcommand{\calA}{\mathcal{A}}

\begin{document}

\title{Optical Ceiling on Polaron Mobility in Disordered Molecular Semiconductors:\\ Lessons from Eumelanin}

\author{Alvan Hunt}
\altaffiliation{Also at Melsa Biotech, Navan, Co.~Meath, Ireland.}
\affiliation{Independent researcher, Dublin, Ireland}

\date{\today}

\begin{abstract}
Polaron hopping in disordered molecular semiconductors such as eumelanin is set by energetic disorder $\sigma$ and reorganization energy $\lambda$. We show that the reduced polaron absorption band encodes this transport problem. Its value at zero energy fixes the mean hop rate, which bounds the mobility in any dimension. For Gaussian disorder and classical vibrations its centroid is $\Eop=\lambda+(1-c)\sigma^2/\kB T$, its variance obeys the Hush relation, and the bound is the Marcus rate at $\lambda\to\Eop$; a 1D lower bound vanishes at $\kB T^*=(1-c)\sigma^2/\lambda$, read from optics.
\end{abstract}

\maketitle

Eumelanin, the brown-black pigment of skin, hair and the substantia nigra, is also one of the oldest known organic semiconductors~\cite{McGinness1974,Meredith2006}. Its building blocks, oxidized indolequinone oligomers, stack into disordered $\pi$ aggregates, and the material shows the hallmarks of an amorphous organic semiconductor in exaggerated form: a broadband, featureless absorption from the ultraviolet to the near infrared~\cite{Tran2006}, strong chemical and conformational disorder, and hydration-dependent mixed electronic and protonic conduction in which the comproportionation equilibrium $\mathrm{H_2Q+Q+2H_2O}\rightleftharpoons\mathrm{2SQ^-+2H_3O^+}$ sets the density of semiquinone carriers~\cite{Mostert2012,Motovilov2019}. This makes eumelanin a natural paradigm for a question that remains open across the field: how energetic disorder and polaronic relaxation combine to set the mobility.

In the standard description, a localized carrier hops between sites with Gaussian distributed energies of width $\sigma$~\cite{Bassler1993}, at Marcus rates set by a reorganization energy $\lambda$~\cite{Marcus1956,Holstein1959}. The low-field mobility is written $\mu\propto\exp[-E_a/\kB T-C(\sigma/\kB T)^2]$, and the coefficient $C$, which weights disorder against the polaron, has been disputed for two decades, with values from $1/8$ to $3/4$ depending on dimension, coordination and $\lambda/\sigma$~\cite{Fishchuk2003,Cottaar2011,Fishchuk2013,Rodin2015,SekiWojcik2016,Saxena2021}. Transport data alone cannot settle the issue, because the temperature dependence of the conductivity mixes $\lambda$, $\sigma$ and, in chemically equilibrated systems such as eumelanin, the activation of the carrier density itself. The polaron's intervalence absorption band, which contains $\lambda$ and $\sigma$ separately, is analyzed with Marcus--Hush theory for isolated pairs~\cite{Hush1967,Brunschwig2002}; its link to transport rests on the small-polaron theory of ordered lattices~\cite{Eagles1963,Reik1963,ReikHeese1967,BottgerBryksin1985}, and the Austin--Mott relation $\Eop=4E_a$ between band peak and hopping barrier is extended to disorder only through a single empirical disorder energy~\cite{AustinMott1969}, yet it is still used to read hopping barriers from polaron absorption~\cite{Tian2026}.

Here we show that the polaron band of thermalized carriers encodes the disordered transport problem through three exact statements. (i) For stationary Gaussian disorder with nearest-neighbor correlation $c$, the centroid of the reduced band is $\Eop=\lambda+(1-c)\sigma^2/\kB T$ and its variance obeys the Hush relation exactly, so thermalized disorder is invisible to the Hush test, whereas high-frequency vibrations and a spread of reorganization energies produce a positive excess, whose temperature slope approximately isolates the vibrational part. (ii) The thermally averaged hop rate equals the band evaluated at zero energy, so the classic Wiener bound of network theory becomes a rigorous optical upper bound on the diffusion coefficient in any dimension; for classical vibrations and Gaussian disorder it is the disorder-free Marcus rate with $\lambda\to\Eop$. (iii) The matching lower bound, the exact one-dimensional result~\cite{SekiTachiya2001}, vanishes at $\kB T^*=(1-c)\sigma^2/\lambda$, which is the ratio of slope to intercept of $\Eop$ versus $1/T$. For short-range-correlated disorder, quasi-one-dimensional transport below $T^*$ is expected to be dispersive with exponent $\alpha=T/T^*$, and quantum vibrations smear the threshold into a crossover constrained by the same spectrum.

\textit{Model.} We consider one excess charge (a semiquinone-type polaron in eumelanin, or a hole in a $\pi$ stack) described by the Holstein Hamiltonian
\begin{align}
H={}&\sum_i\varepsilon_i n_i+J\sum_{\langle ij\rangle}(c_i^\dagger c_j+\mathrm{h.c.})\nonumber\\
&+\sum_q\hbar\omega_q b_q^\dagger b_q+\sum_{i,q}g_{iq}n_i(b_q+b_q^\dagger),
\label{eq:H}
\end{align}
with $n_i=c_i^\dagger c_i$ and stationary Gaussian site energies, $\avg{\varepsilon_i}=0$, $\avg{\varepsilon_i^2}=\sigma^2$, and $\avg{\varepsilon_i\varepsilon_j}=c\sigma^2$ for nearest neighbors. In the nonadiabatic limit $J\ll\lambda,\kB T$ with classical vibrations, hops occur at the golden-rule rates
\begin{equation}
w_{ij}=\frac{2\pi}{\hbar}|J|^2F_{ij}(0),\quad
F_{ij}(E)=\frac{e^{-(E-\lambda-\Delta_{ij})^2/4\lambda\kB T}}{\sqrt{4\pi\lambda\kB T}},
\label{eq:rate}
\end{equation}
where $\Delta_{ij}=\varepsilon_j-\varepsilon_i$. The Franck--Condon weighted density $F_{ij}(E)$ describes transfer of the charge from $i$ to $j$ while absorbing energy $E$; the same function controls the optical transition at $E=\hbar\omega$, and this is the bridge we exploit. Detailed balance makes $C_{ij}\equiv e^{-\beta\varepsilon_i}w_{ij}$ symmetric,
\begin{equation}
C_{ij}=\nu\exp\Bigl\{-\beta\Bigl[\frac{\varepsilon_i+\varepsilon_j}{2}+\frac{(\varepsilon_i-\varepsilon_j)^2}{4\lambda}+\frac{\lambda}{4}\Bigr]\Bigr\},
\label{eq:C}
\end{equation}
with $\beta=1/\kB T$ and $\nu=(|J|^2/\hbar)(\pi/\lambda\kB T)^{1/2}$. In reduced densities $u_i=P_ie^{\beta\varepsilon_i}$, the master equation $\dot P_i=\sum_j(w_{ji}P_j-w_{ij}P_i)$ is Kirchhoff's law for a network with bond conductances $C_{ij}$ and site capacities $e^{-\beta\varepsilon_i}$. For stationary ergodic disorder the macroscopic diffusion coefficient follows by homogenization~\cite{Kirkpatrick1973,HausKehr1987},
\begin{equation}
D=a^2\,\sigma_{\mathrm{eff}}/\avg{e^{-\beta\varepsilon}},
\label{eq:D}
\end{equation}
where $\sigma_{\mathrm{eff}}$ is the effective bond conductance of $\{C_{ij}\}$ and $a$ the lattice spacing; the mobility is $\mu=eD/\kB T$.

\textit{Polaron band.} A thermalized carrier on site $i$ absorbs a photon by vertical transfer to a neighbor $j$. In the Condon approximation the transition dipole is $e r_{ij}J/\hbar\omega$ (Mulliken--Hush~\cite{Brunschwig2002}), so the reduced line shape
\begin{equation}
\calA(E)\equiv\avg{\pi_iF_{ij}(E)}\propto\frac{E\,\alpha(E)}{1-e^{-\beta E}}
\label{eq:A}
\end{equation}
is measured by the polaron-induced absorption coefficient $\alpha$ at $E=\hbar\omega$. Here $\pi_i=e^{-\beta\varepsilon_i}/\avg{e^{-\beta\varepsilon}}$, and the average runs over sites and neighbor directions. Two properties hold for any disorder law. Exchanging $i\leftrightarrow j$ with detailed balance gives the McCumber (Kennard--Stepanov) symmetry $\calA(-E)=e^{-\beta E}\calA(E)$~\cite{McCumber1964}. It also completes the measured band, which is accessible only at $E>0$: moments follow from $\calA$ at $E>0$ plus its reflection $e^{-\beta E}\calA(E)$ at $-E$, normalized to unit area; for a Gaussian band the reflected part carries less than $1\%$ of the weight once $\Eop>11\,\kB T$. And Eq.~\eqref{eq:rate} gives directly
\begin{equation}
\avg{\pi_iw_{ij}}=\frac{2\pi}{\hbar}|J|^2\calA(0):
\label{eq:bridge}
\end{equation}
the thermally averaged hop rate is the band evaluated at zero energy.

For Gaussian disorder, the Boltzmann weight shifts the occupied site to $\varepsilon_i\sim N(-\beta\sigma^2,\sigma^2)$, while $\varepsilon_j|\varepsilon_i\sim N(c\varepsilon_i,\sigma^2(1-c^2))$, so that $\Delta_{ij}\sim N((1-c)\beta\sigma^2,2(1-c)\sigma^2)$. Convolving with Eq.~\eqref{eq:rate} yields a Gaussian band whose centroid (first moment) $\Eop$ and variance $W^2$ are
\begin{align}
\Eop&=\lambda+\frac{(1-c)\sigma^2}{\kB T},\nonumber\\
W^2&=2\lambda\kB T+2(1-c)\sigma^2=2\kB T\,\Eop.
\label{eq:Eop}
\end{align}
The second equality is the Hush relation, $\mathrm{FWHM}^2=16\ln2\,\kB T\Eop$~\cite{Hush1967}, here exact for arbitrarily strong thermalized disorder: disorder shifts and broadens the band in precisely the ratio that detailed balance requires. In molecular mixed-valence systems, departures from the Hush width are used to diagnose electronic delocalization~\cite{Brunschwig2002}; here thermalized site disorder, however strong, produces none. An analogous thermalization shift, $-\sigma^2/\kB T$, is known for emission from charge-transfer states at organic heterojunctions~\cite{Burke2015}, but there only the initial state is thermally selected and the Hush relation is not restored; for intervalence transfer both sites are disordered and it holds exactly. An excess width therefore signals physics beyond the classical homogeneous model. Adding one mode of energy $\hbar\Omega$ and Huang--Rhys factor $S$ to Eq.~\eqref{eq:rate}~\cite{BixonJortner1999} makes the band non-Gaussian, so its peak no longer coincides with its centroid, but the cumulants still add exactly: the centroid becomes $\Eop=\lambda+S\hbar\Omega+(1-c)\sigma^2/\kB T$, with unchanged slope in $1/T$, and
\begin{equation}
W^2-2\kB T\Eop=S\hbar\Omega\Bigl[\hbar\Omega\coth\frac{\hbar\Omega}{2\kB T}-2\kB T\Bigr]\ge0,
\label{eq:excess}
\end{equation}
which vanishes classically, falls as $S(\hbar\Omega)^3/6\kB T$ at high temperature and saturates at $S(\hbar\Omega)^2$ at low temperature. A spread of reorganization energies, independent of the site energies, adds a temperature-independent $\mathrm{Var}(\lambda)$ to the right-hand side. At accessible temperatures, $\kB T\ll\hbar\Omega$, the excess is nearly linear, $S\hbar\Omega(\hbar\Omega-2\kB T)+\mathrm{Var}(\lambda)$, and its slope approximates $S\hbar\Omega$, the correction that turns the centroid intercept into $\lambda$. Residual curvature biases a two-point slope low (by $7\%$ over 250--400~K for $\hbar\Omega=0.18$~eV); fitting Eq.~\eqref{eq:excess} with $\hbar\Omega$ fixed from Raman spectra removes this bias, and an independent mode energy or resolved sidebands are needed anyway to separate $S$, $\hbar\Omega$ and $\mathrm{Var}(\lambda)$. Evaluating Eq.~\eqref{eq:bridge} on the Gaussian band of Eq.~\eqref{eq:Eop} gives
\begin{equation}
\avg{\pi_iw_{ij}}=\frac{|J|^2}{\hbar}\sqrt{\frac{\pi}{\Eop\kB T}}\,e^{-\Eop/4\kB T}\equiv k_M(\Eop),
\label{eq:kM}
\end{equation}
the disorder-free Marcus rate $k_M$ with $\lambda$ replaced by $\Eop$.

\textit{Theorem.} For nearest-neighbor hopping on a hypercubic lattice in any dimension, with stationary, ergodic and statistically isotropic disorder,
\begin{equation}
\frac{a^2}{\avg{e^{-\beta\varepsilon}}\avg{C^{-1}}}\;\le\;D\;\le\;a^2\avg{\pi_iw_{ij}}=\frac{2\pi a^2}{\hbar}|J|^2\calA(0).
\label{eq:bounds}
\end{equation}
\textit{Proof.} The conductance $\sigma_{\mathrm{eff}}$ of Eq.~\eqref{eq:D} obeys the Dirichlet and Thomson variational principles~\cite{DoyleSnell1984}. The trial potential $u=x$, with a unit drop across every bond along $x$ and none across the others, gives $\sigma_{\mathrm{eff}}\le\avg{C}$. A unit current flowing straight along each row of $x$ bonds is divergence free and gives $\sigma_{\mathrm{eff}}^{-1}\le\avg{C^{-1}}$. Dividing by $\avg{e^{-\beta\varepsilon}}$ and using $\avg{C}/\avg{e^{-\beta\varepsilon}}=\avg{\pi_iw_{ij}}$ with Eq.~\eqref{eq:bridge} completes the proof. $\square$ These are the classic Wiener bounds~\cite{Wiener1912}, the harmonic and arithmetic means of the bond conductances; what is new is the identification of the upper one with the band at zero energy.

The upper bound needs only golden-rule rates in the Condon approximation, so it holds with quantum modes and non-Gaussian disorder, provided $\calA(0)$ is taken from the measured band. In practice $\calA(0)$ lies in the deep low-energy tail; for classical vibrations and Gaussian disorder it follows from the centroid through Eq.~\eqref{eq:kM}, and otherwise from a fit of the line shape. It is saturated in mean field (coordination $z\to\infty$), where for Gaussian disorder it gives $C=(1-c)/4$, the value previously found as an approximate limit of effective-medium theory~\cite{Rodin2015,SekiWojcik2016}; it is the dc counterpart of the familiar rise of hopping conductivity toward its high-frequency limit~\cite{DyreSchroder2000}. For classical vibrations and Gaussian disorder, Eq.~\eqref{eq:kM} turns it into an optical Austin--Mott inequality,
\begin{equation}
\mu(T)\le\frac{ea^2}{\kB T}\,k_M\bigl(\Eop(T)\bigr),
\label{eq:AM}
\end{equation}
which for $\sigma=0$ reduces to the Austin--Mott equality $\Eop=4E_a$, with $E_a=\lambda/4$ the Marcus barrier rather than the full apparent activation energy, which also contains the prefactor. Disorder turns the Austin--Mott relation into a one-sided bound in which disorder enters only through the measured $\Eop$; the prefactor $a^2|J|^2$ follows, up to geometric factors, from the integrated intensity of the same band (Mulliken--Hush), given the carrier density.

\textit{Bottleneck temperature.} In one dimension the bonds are in series, $\sigma_{\mathrm{eff}}=\avg{C^{-1}}^{-1}$, and the lower bound in Eq.~\eqref{eq:bounds} is exact~\cite{Derrida1983}. The variables $s=(\varepsilon_i+\varepsilon_{i+1})/2$ and $d=\varepsilon_{i+1}-\varepsilon_i$ are independent Gaussians with variances $\sigma^2(1+c)/2$ and $2\sigma^2(1-c)$, the averages are elementary, and
\begin{align}
D_{1\mathrm{D}}&=a^2\nu\,\exp\Bigl[-\frac{\lambda}{4\kB T}-\frac{(3+c)\,\sigma^2}{4(\kB T)^2}\Bigr]\sqrt{1-\frac{T^*}{T}},\nonumber\\
\kB T^*&=\frac{(1-c)\,\sigma^2}{\lambda},
\label{eq:D1D}
\end{align}
for $T>T^*$, while $\avg{C^{-1}}$ diverges and $D_{1\mathrm{D}}=0$ for $T\le T^*$. This is the result of Seki and Tachiya from mean first-passage times [Eqs.~(6.7) and (6.13) of Ref.~\cite{SekiTachiya2001}]; the factor $(1-T^*/T)^{1/2}$ is absent from the large-$\lambda$ form usually quoted~\cite{SekiWojcik2016}. Equation~\eqref{eq:D1D} reduces to the ordered Marcus result for $\sigma\to0$, to the large-$\lambda$ form for $\lambda\to\infty$, and to Zwanzig's result $D\propto e^{-\beta^2\sigma^2}$ for a smooth Gaussian landscape as $c\to1^-$, with the correlation length large but finite~\cite{Zwanzig1988}.

The zero of Eq.~\eqref{eq:D1D} is transparent in network language. For $T<T^*$ the average $\avg{e^{\beta d^2/4\lambda}}$ diverges because of rare bonds with an energy step $|d|\gg\lambda$. Uphill, such a hop is Boltzmann suppressed; downhill, it lies deep in the Marcus inverted region. A large energy step is therefore a barrier in both directions, which Miller--Abrahams rates, with fast downhill hops, cannot produce. A large-deviation estimate gives the tail of the bond resistance,
\begin{equation}
\mathrm{Prob}\bigl(C^{-1}>R\bigr)\sim R^{-\alpha}/(\ln R)^{1/2},\qquad \alpha=T/T^*.
\label{eq:tail}
\end{equation}
For short-range-correlated disorder, where rare steps are effectively independent, the chain at $T<T^*$ then falls in the universality class of the random-barrier model with broad barrier distribution~\cite{Alexander1981,BouchaudGeorges1990}: the resistance of $N$ bonds grows as $N^{1/\alpha}$ up to logarithmic corrections, and $\avg{x^2}\propto t^{2\alpha/(1+\alpha)}$. Under bias we therefore expect, pending a finite-field calculation, transit times across a film of thickness $L$ scaling as $L^{1/\alpha}$, dispersive time-of-flight transients with Scher--Montroll parameter $\alpha=T/T^*$~\cite{ScherMontroll1975}, and an apparent mobility falling with thickness as $L^{1-1/\alpha}$. Seki and Tachiya attributed the vanishing of Eq.~\eqref{eq:D1D} to the infinite-chain limit~\cite{SekiTachiya2001}; Eq.~\eqref{eq:tail} shows that it is a threshold with observable finite-size consequences.

\begin{figure}[t]
\begin{tikzpicture}
\begin{axis}[
  scale only axis, width=0.80\columnwidth, height=0.44\columnwidth,
  xmin=0, xmax=3, ymin=-8.3, ymax=0.4,
  xlabel={$\sigma/\kB T$}, ylabel={$\ln(D/D_0)$},
  label style={font=\small}, tick label style={font=\footnotesize},
  legend style={font=\scriptsize, draw=none, fill=none, at={(0.02,0.02)}, anchor=south west},
  legend cell align=left]
\addplot[thick, domain=0:3, samples=120] {-0.5*ln(1+x^2/10) - x^2/4};
\addlegendentry{optical upper bound, $a^2k_M(\Eop)$}
\addplot[thick, dashed, domain=0:3, samples=160] {-0.75*x^2 + 0.5*ln(1-x^2/10)};
\addlegendentry{exact 1D, lower bound}
\addplot[only marks, mark=square*, mark size=1.4pt] coordinates
  {(0.5,-0.1231) (1.0,-0.4912) (1.5,-1.0983) (2.0,-1.9731) (2.5,-3.1003) (2.8,-3.9109)};
\addlegendentry{network, $d=2$}
\addplot[only marks, mark=*, mark size=1.5pt] coordinates
  {(0.5,-0.1021) (1.0,-0.4093) (1.5,-0.9132) (2.0,-1.6524) (2.5,-2.6142) (2.8,-3.3158)};
\addlegendentry{network, $d=3$}
\node[font=\small] at (rel axis cs:0.94,0.90) {(a)};
\end{axis}
\end{tikzpicture}

\begin{tikzpicture}
\begin{axis}[
  scale only axis, width=0.80\columnwidth, height=0.44\columnwidth,
  xmin=0.4, xmax=1.55, ymin=-5.6, ymax=0.3,
  xlabel={$T^*/T$}, ylabel={$\ln(D/D_{3/4})$},
  label style={font=\small}, tick label style={font=\footnotesize},
  legend style={font=\scriptsize, draw=none, fill=none, at={(0.02,0.02)}, anchor=south west},
  legend cell align=left]
\addplot[thick, domain=0.4:0.99999, samples=600] {0.5*ln(1-x)};
\addlegendentry{$S=0$ (classical)}
\addplot[thick, dashed] coordinates
  {(0.40,-0.243) (0.45,-0.284) (0.50,-0.329) (0.55,-0.379) (0.60,-0.434) (0.65,-0.495) (0.70,-0.563) (0.75,-0.641) (0.80,-0.731) (0.85,-0.835) (0.90,-0.955) (0.95,-1.098) (1.00,-1.265) (1.05,-1.463) (1.10,-1.695) (1.15,-1.964) (1.20,-2.273) (1.25,-2.620) (1.30,-3.006) (1.35,-3.427) (1.40,-3.880) (1.45,-4.361) (1.50,-4.867) (1.55,-5.396)};
\addlegendentry{$S=0.1$}
\addplot[thick, dashdotted] coordinates
  {(0.40,-0.221) (0.45,-0.259) (0.50,-0.300) (0.55,-0.345) (0.60,-0.394) (0.65,-0.449) (0.70,-0.509) (0.75,-0.576) (0.80,-0.651) (0.85,-0.735) (0.90,-0.830) (0.95,-0.938) (1.00,-1.060) (1.05,-1.199) (1.10,-1.357) (1.15,-1.536) (1.20,-1.738) (1.25,-1.965) (1.30,-2.217) (1.35,-2.496) (1.40,-2.802) (1.45,-3.132) (1.50,-3.488) (1.55,-3.867)};
\addlegendentry{$S=0.3$}
\addplot[thick, dotted] coordinates
  {(0.40,-0.167) (0.45,-0.196) (0.50,-0.229) (0.55,-0.265) (0.60,-0.304) (0.65,-0.346) (0.70,-0.393) (0.75,-0.445) (0.80,-0.502) (0.85,-0.564) (0.90,-0.633) (0.95,-0.709) (1.00,-0.793) (1.05,-0.886) (1.10,-0.989) (1.15,-1.103) (1.20,-1.229) (1.25,-1.369) (1.30,-1.523) (1.35,-1.694) (1.40,-1.882) (1.45,-2.087) (1.50,-2.311) (1.55,-2.555)};
\addlegendentry{$S=1$}
\draw[gray, very thin] (axis cs:1,-5.6) -- (axis cs:1,0.3);
\node[font=\small] at (rel axis cs:0.94,0.90) {(b)};
\end{axis}
\end{tikzpicture}
\caption{(a)~Disorder suppression of the diffusion coefficient at $\lambda=10\,\kB T$, $c=0$, relative to the ordered value $D_0=a^2k_M(\lambda)$. Solid line: optical upper bound of Eq.~\eqref{eq:bounds}; dashed line: exact 1D lower bound, Eq.~\eqref{eq:D1D}; symbols: exact network solutions on $160^2$ (squares) and $32^3$ (circles) lattices, averaged over 6 and 4 disorder realizations (statistical errors below 0.01). (b)~Exact 1D diffusion coefficient relative to the large-$\lambda$ form $D_{3/4}=D_0e^{-3\sigma^2/4(\kB T)^2}$, where $D_0$ is the ordered value for the same rates, for classical Marcus rates and for Marcus--Levich--Jortner rates with one mode $\hbar\Omega=0.18$~eV of Huang--Rhys factor $S$ ($\sigma=0.10$~eV, $\lambda=0.30$~eV, $c=0$, $T^*=387$~K). The vertical line marks $T=T^*$.}
\label{fig1}
\end{figure}
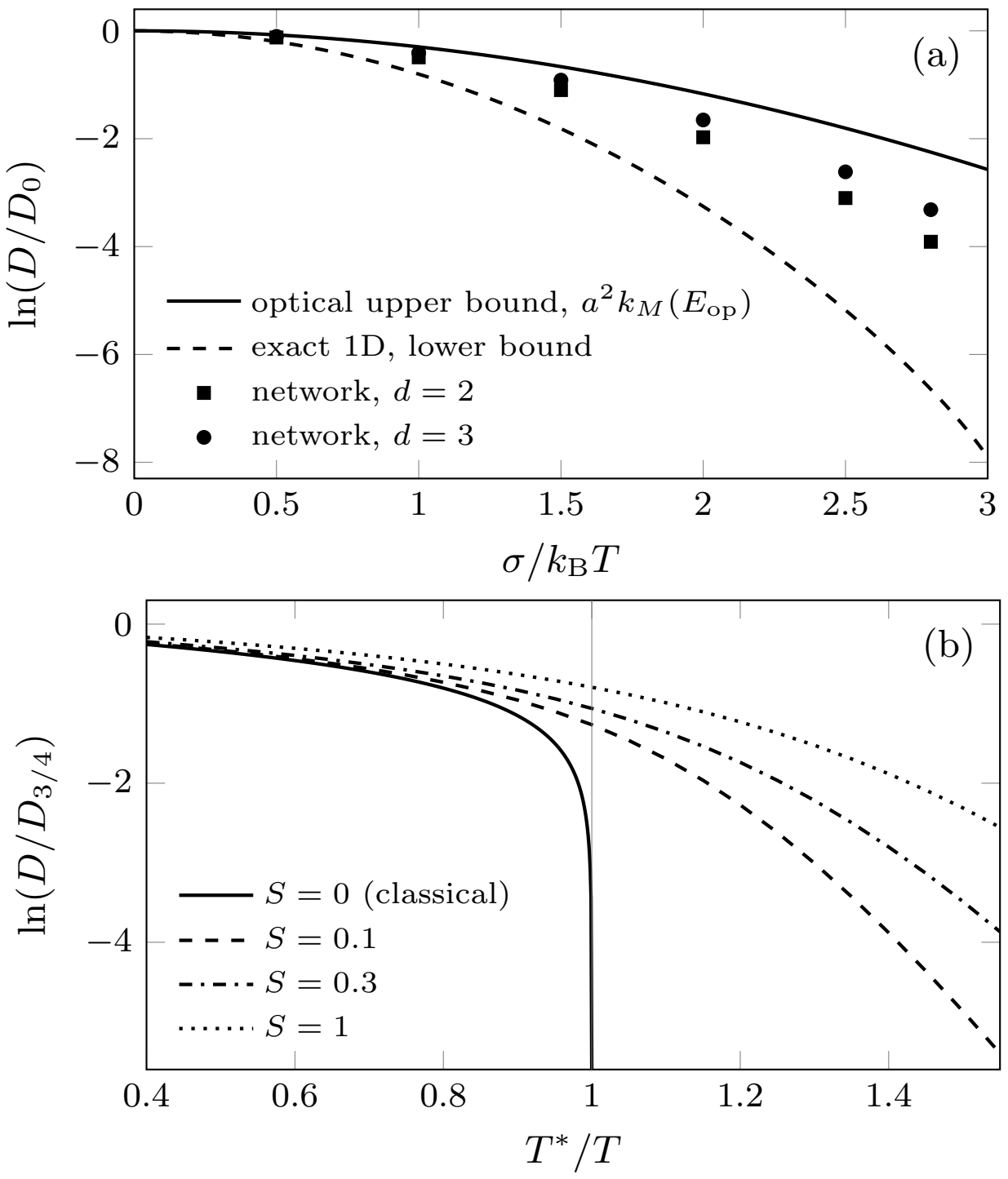

Crucially, $T^*$ is fixed by optics. If $\lambda$, $\sigma$ and $c$ are temperature independent, Eq.~\eqref{eq:Eop} makes $\Eop$ linear in $1/T$ with intercept $\lambda$ and slope $(1-c)\sigma^2/\kB$, so
\begin{equation}
T^*=\frac{\text{slope}}{\text{intercept}},\qquad\text{equivalently}\qquad \Eop(T^*)=2\lambda .
\label{eq:Tstar}
\end{equation}
The same combination $(1-c)\sigma^2$ enters both quantities, so short-range correlations do not spoil the readout. With quantum modes the slope is unchanged and the intercept becomes $\lambda+S\hbar\Omega$, corrected via Eq.~\eqref{eq:excess}. Since $\delta T^*/T^*\simeq-\delta\lambda/\lambda$, the uncertainty of this correction carries directly into $T^*$: at most $4\%$ for $S\le1$ with a two-point slope and the parameters of Fig.~\ref{fig1}(b), and more when $\lambda$ is small. $T^*$ then remains the reference scale of the underlying classical bottleneck. With a spread of $\lambda$ the intercept gives a coupling-weighted mean of $\lambda$, and hence a lower estimate of the threshold. In $d>1$ the lower bound also vanishes at $T^*$; transport then proceeds through detours around the steps, bracketed by Eq.~\eqref{eq:bounds}.

\textit{Quantum smearing and robustness.} Real $\pi$-conjugated sites carry high-frequency accepting modes (C=C and C=O stretches, $\hbar\Omega\approx0.18$~eV). For a downhill gap $|d|\gg\lambda$, the channel with $n\approx(|d|-\lambda)/\hbar\Omega$ quanta leaves a classical mismatch below $\hbar\Omega/2$, and its thermal weight, $\propto[S(\bar n+1)]^n/n!$ with $\bar n$ the mode occupation, makes $-\ln w$ grow only as $|d|\ln|d|$, the energy-gap law~\cite{EnglmanJortner1970}. This is slower than the Gaussian cost $d^2/4(1-c)\sigma^2$ of the step, so for an ideal harmonic mode at finite temperature $\avg{C^{-1}}$ stays finite for any $S>0$ and the threshold becomes a crossover. Figure~\ref{fig1}(b) shows the exact 1D diffusion coefficient with Marcus--Levich--Jortner rates, including thermal mode populations so that detailed balance holds. The classical curve follows $\frac12\ln(1-T^*/T)$ and terminates at $T^*$; for $S=0.1$ to $1$ the diffusion coefficient stays finite but falls increasingly below the large-$\lambda$ form on cooling, by factors of 10 to 130 at $T=T^*/1.5$. The degree of smearing is set by $S$ and $\hbar\Omega$, which the band variance constrains through Eq.~\eqref{eq:excess} given an independent mode energy. Two further effects soften the threshold. Hopping beyond nearest neighbors lets carriers bypass single steps; for independent site energies and comparable couplings, blocking range-$m$ hopping requires $m$ low sites followed by $m$ high ones, which suggests a threshold near $T^*/m$. And a chain of $N$ sites contains steps only up to $\sim2\sigma[(1-c)\ln N]^{1/2}$. We verified Eqs.~\eqref{eq:Eop}, \eqref{eq:kM} and \eqref{eq:D1D} by Monte Carlo sampling (agreement within statistical error, at most 0.2\%), Eq.~\eqref{eq:excess} with and without a spread of $\lambda$ by exact sampling of the band, the convergence of Fig.~\ref{fig1}(b) in the vibrational and step cutoffs (to $10^{-4}$), the predicted growth of the chain resistance as $N^{\max(1,1/\alpha)}$, approached slowly as the logarithm in Eq.~\eqref{eq:tail} implies (this supports, but does not test, the finite-field predictions), and Eq.~\eqref{eq:bounds} by solving the network exactly in two and three dimensions [Fig.~\ref{fig1}(a)]; the fitted coefficients $C=0.50$ ($d=2$) and $0.42$ ($d=3$) agree with kinetic Monte Carlo~\cite{SekiWojcik2016}.

\textit{Eumelanin.} Three features make eumelanin a natural test bed. First, comproportionation is a chemical doping knob: small hydration steps change the semiquinone density strongly~\cite{Mostert2012}. To the extent that they leave the band shape nearly unchanged, an assumption to be tested, hydration-difference spectra isolate the reduced polaron band from the much stronger chromophore background, and a temperature series of its centroid and variance yields $(1-c)\sigma^2$, $\lambda$ and $T^*$ from Eqs.~\eqref{eq:Eop}, \eqref{eq:excess} and \eqref{eq:Tstar}, and constrains $S$, $\hbar\Omega$ and the spread of $\lambda$. Because $\Eop$ and $W$ do not depend on the carrier density, this separates the mobility from the thermally activated density, which dc measurements cannot do. The hopping-regime activation energies from broadband dielectric spectroscopy, $0.43$~eV dry and $0.52$~eV hydrated~\cite{Motovilov2019}, illustrate the ambiguity: Eq.~\eqref{eq:D1D} with $\lambda=0.4$~eV and $\sigma=0.08$~eV ($T^*\approx186$~K) gives a mobility activation energy $-d\ln\mu/d\beta=0.45$~eV at 300~K with no density activation at all, yet the data are equally compatible with a strongly activated density. Second, over larger hydration ranges absorbed water should raise the outer-sphere reorganization energy; at fixed $\sigma$ this lowers $T^*=(1-c)\sigma^2/\kB\lambda$, so hydration would pull the material away from the bottleneck regime, a prediction specific to the polaron picture. Third, eumelanin is polydisperse, so oligomer size spreads $\lambda$ and $J$ from bond to bond. For independent bond reorganization energies $\lambda_b$ bounded above and away from zero, and couplings bounded away from zero, the 1D threshold is set by the most delocalized bonds, $\kB T^*=(1-c)\sigma^2/\lambda_{\min}$ with $\lambda_{\min}$ the essential infimum, so polydispersity raises $T^*$; a heavy upper tail of $\lambda_b$, or couplings approaching zero, would add bottlenecks of their own. The optical intercept is the coupling-weighted mean $\avg{|J_b|^2\lambda_b}/\avg{|J_b|^2}\ge\lambda_{\min}$ (the same weighting enters $\mathrm{Var}(\lambda)$ in Eq.~\eqref{eq:excess}), so the readout gives a lower estimate. For bond-dependent couplings $J_b$ the Condon dipole weights each transition by $|J_b|^2$, so the measured band supplies exactly the combination $\avg{\pi_i|J_{ij}|^2F_{ij}(0)}$ that enters $\avg{\pi_iw_{ij}}$; the upper bound of Eq.~\eqref{eq:bounds} therefore still holds when read from the measured band, although Eq.~\eqref{eq:kM} becomes approximate. Individual protomolecule stacks are only a few layers deep, so the one-dimensional bottleneck concerns long percolation paths through aggregates, while Eq.~\eqref{eq:bounds} applies to the bulk regardless. We are not aware of an assignment of the semiquinone polaron band in eumelanin; resolving it is the central experimental task.

\textit{Outlook.} The results are not specific to eumelanin. Any disordered polaron conductor whose charged state has a resolvable intervalence band, from conjugated polymers probed by charge-modulation spectroscopy and donor or acceptor phases in organic photovoltaic blends to columnar discotics, DNA and protein nanowires, can be placed against the ceiling of Eq.~\eqref{eq:bounds}, which takes the Marcus form of Eq.~\eqref{eq:AM} when vibrations are classical, and quasi-one-dimensional systems can be tested for the threshold of Eqs.~\eqref{eq:tail} and \eqref{eq:Tstar}. For bioelectronics this turns an empirical figure of merit into a spectroscopic design target: in the classical regime, raising the mobility ceiling means lowering $\Eop$, that is, reducing $\lambda$ or the effective disorder $(1-c)\sigma^2$, while the excess width of Eq.~\eqref{eq:excess} shows when vibronic structure must be included. For synthetic melanins and polydopamine films, where polymerization conditions tune oligomer size and stacking, the same spectral series quantifies how processing moves a material relative to its ceiling and its bottleneck temperature. The main limitations are the dilute-carrier, nearest-neighbor and nonadiabatic assumptions; state filling at high doping modifies the weights $\pi_i$, and adiabatic transfer modifies Eq.~\eqref{eq:rate}, both natural extensions.

\begin{acknowledgments}
The author is the founder and chief executive of Melsa Biotech, a producer of eumelanin; this work was carried out independently of the company. The author conceived the project, set its focus on eumelanin and its staged structure (ideation, derivation, manuscript), made the key decisions at each stage, and obtained the prior work of Ref.~\cite{SekiTachiya2001} whose examination led to the final framing. Candidate research directions, analytical derivations, numerical checks, literature searches and the first draft were generated with the AI system Claude (model Claude Opus 5.5, Anthropic) under the author's direction. A second AI system, ChatGPT 6 Astra (OpenAI), reviewed the manuscript over several rounds, re-derived the main results and checked them with independently written code (direct quadrature, a separate network solver and an alternative construction of the quantum rates); the resulting corrections were incorporated at the author's direction. All analytical results were checked against independent numerical calculations (Monte Carlo sampling and exact network solutions), and references were checked against publisher, library and PubMed records. The author takes full responsibility for the content. Python code that reproduces every numerical result and both panels of Fig.~\ref{fig1} from fixed random seeds is provided as ancillary files.
\end{acknowledgments}

\end{document}